# Magnetic ground state discrimination of a Polyradical Nanographene using Nickelocene-Functionalized Tips

Diego Soler-Polo,[1,*] Oleksandr Stetsovych,[1] Manish Kumar,[1] Benjamin Lowe,[1] Ana Barragán,[2] Zhiqiang Gao,[3] Andrés Pinar Solé,[1] Hao Zhao,[3] Elena Pérez-Elvira,[2] Goudappagouda,[3] David Écija,[1] Akimitsu Narita,[3,*] Pavel Jelínek[1,*] and José I. Urgel[2,*]

[1]Institute of Physics of the Czech Academy of Science, CZ-16253 Praha, Czech Republic. [2]IMDEA Nanoscience, C/ Faraday 9, Campus de Cantoblanco, 28049 Madrid, Spain. [3]Organic and Carbon Nanomaterials Unit, Okinawa Institute of Science and Technology Graduate University 1919-1 Tancha, Onna-son, Kunigami-gun, Okinawa 904-0495, Japan. ✉e-mail: soler@fzu.cz; akimitsu.narita@oist.jp; jelinekp@fzu.cz and jose-ignacio.urgel@imdea.org

## Abstract

Molecular magnets are a promising class of materials with exciting properties and applications. However, a profound understanding and application of such materials depends on the accurate detection of their electronic and magnetic properties. Despite the availability of experimental techniques that can sense the magnetic signal, the exact determination of the spin ground states and spatial distribution of exchange interaction of strongly correlated single-molecule magnets remains challenging. Here, we demonstrate that scanning probe microscopy with a nickelocene-functionalized probe can distinguish between nearly degenerate multireference ground states of single-molecule π-magnets and map their spatial distribution of the exchange interaction. This method expands the already outstanding imaging capabilities of scanning probe microscopy for characterizing the chemical and electronic structures of individual molecules, paving the way for the study of strongly correlated molecular magnets with unprecedented spatial resolution.

## Main

Single-molecule magnets represent an interesting class of materials with great application potential. One of the key factors for their future use in optoelectronics and spintronics depends on the ability to characterize their electronic and magnetic properties at the single-molecule level.[1] Traditional molecular magnets are based on metal-organic compounds where the magnetic moment originates from strongly localized d- and f-electrons on metal centers.[2,3] On the other hand, the magnetism of organic carbon-based radicals, the so-called π-magnetism,[4,5] is mainly associated with highly delocalized unpaired π-electrons.[6] Thus, a deeper understanding of the π-magnets depends not only on determining the correct magnetic ground state of a given molecule but also on our ability to resolve the spatial distribution of the inhomogeneous magnetic signal.

Traditional techniques, such as Electron Paramagnetic Resonance (EPR) and Superconducting Quantum Interference Device (SQUID) magnetometry, are extensively employed to study the magnetic properties of many organic radicals.[7,8] While EPR provides valuable insights into the electronic environments of unpaired electrons, SQUID magnetometry is capable of measuring the total magnetic moment of bulk samples. These techniques can be complemented by Electron-nuclear double resonance spectroscopy (ENDOR),[9] Variable Magnetic Field Scanning (VMS), and Magneto-Optical

Kerr Effect (MOKE), among others, to provide a more comprehensive understanding of the magnetization and other magnetic characteristics of targeted organic radicals. All these methods, though powerful, are primarily designed for the characterization of molecular assemblies and are often limited by low sensitivity for defects or spurious intermolecular interactions. In addition, the unambiguous determination of the ground magnetic state of strongly correlated molecules with nearly degenerate electronic states represents a non-trivial task for all these methods. But more importantly, all these techniques have limited capability to provide detailed information about the spatial distribution of the magnetic signal on a single molecule. The spatial distribution is especially relevant for molecular π-magnets, where the magnetic moment is determined by strongly delocalized unpaired π-electrons, which can create a spatially inhomogeneous exchange interaction.

In recent decades, the characterization of single-molecule π-magnets with tailor-made magnetic ground states[10-12] and the prospect for spintronics and quantum technologies [4,13-15] has been expanded beyond traditional methods to include advanced surface science techniques. This expansion is driven by the emergent concept of on-surface synthesis,[16] which enables the fabrication of otherwise unstable π-magnets on metal surfaces facilitated by the employed ultra-high vacuum (UHV) environment. Moreover, this concept can be naturally complemented with low-temperature UHV scanning probe microscopy (SPM) techniques, which provide valuable insights into the chemical and electronic structure of single molecules with unprecedented spatial resolution.[6] Nevertheless, a reliable determination of their magnetic ground and excited states remains a significant challenge.[9] Typically, the magnetic nature of on-surface synthesized molecules with well-defined magnetic ground states, i.e., open-shell nanographenes (NGs), is inferred from the spectroscopic features acquired with SPM tip. Examples are Kondo-like resonances, Coulomb gaps, or inelastic spin-flip excitations with spatially-resolved magnetic signals.[17-19] These experimental findings are usually reinforced by state-of-the-art multireference calculations that predict the ground state of the studied nanostructures. However, this might not be enough to discriminate the magnetic state or systems with nearly degenerate states. Identifying then the nature of the ground state is hindered by similar spectral features and excitation values below the precision of ab-initio multireference calculations.

Recently, EPR has been combined with Scanning Tunneling Microscopy (STM), integrating the precision of STM with the spin-sensitive capabilities of EPR. The EPR-STM technique is used to image, characterize, and coherently control individual metal atom spins and single coordination complexes on insulator surfaces.[20] EPR-STM combines the unprecedented energy resolution of EPR with the atomic-scale precision of STM.[20] It can be used to image, characterize, and coherently control spins on surfaces. However, its application has thus far been mostly limited to metal atoms[21] or metal-containing molecules[22] adsorbed on MgO bilayers on Ag(100). Note that recently ESR signal was also detected by STM probe functionalized by a single molecule.[23] In addition to the EPR-STM technique, other well-known surface-sensitive techniques like X-ray Magnetic Circular Dichroism (XMCD) and Light Magnetic Dichroism (LMD) are often used to study the magnetic properties of transition metals and heavier elements. Unfortunately, these techniques present difficulties when applied to carbon-based nanomaterials due to the low atomic number of carbon atoms and the weak magnetic response, together with the need for well-defined surface structures.

Another alternative is spin-polarized STM with sharp magnetic tips,[24] which employs the effect of tunnel magnetoresistance to achieve atomic-scale spin contrast. Nevertheless, this method cannot discriminate different molecular spin states. Also, the significant chemical reactivity of atomically sharp magnetic tips hampers stable scanning conditions at close tip-sample distances.

Under this scenario, tip functionalization with a nickelocene (NiCp$_2$) molecule as spin-sensitive probes or magnetic field sensors (with or without the presence of a magnetic field)[25–31] has recently emerged as an innovative approach providing quantitative spin-dependent information.[28] The NiCp$_2$ molecule has total spin S=1, where the spin-orbit coupling causes a split of the triplet state into the in-plane spin ground state (m$_s$=0) and the doubly degenerate spin-up and spin-down excited state (m$_s$=±1). Importantly, the net spin (S = 1) of NiCp$_2$ placed on the metallic tip apex remains preserved from scattering events with itinerant electrons from the metal tip. Thus, inelastic electron tunneling spectroscopy (IETS) acquired with a NiCp$_2$-functionalized probe shows a large enhancement of the inelastic signal at 4 meV as a consequence of the spin excitation from the ground to the first excited state.[25] The presence of the exchange interaction between a NiCp$_2$-probe and a magnetic system on the surface can modify the characteristic IETS signal along SPM tip approach, as shown in the inset of Figure 1. Therefore, the variation of the IETS signal at short tip-sample distances provides valuable information about the exchange interaction as well as the spin states of the inspected system by NiCp$_2$-probe. This strategy was employed to probe the magnetic signal of single atoms,[25] molecules,[26] one-dimensional metal-organic chains[29], and 2D materials.[32]

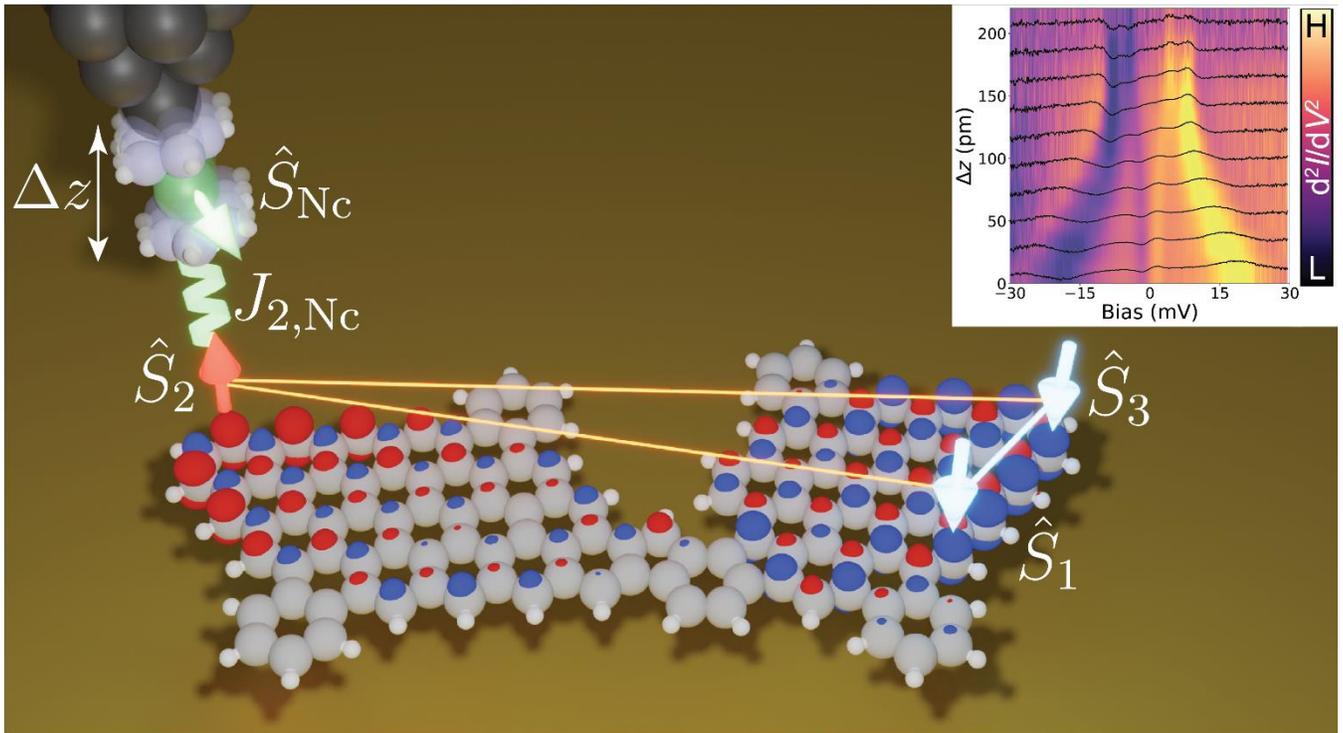

$$\hat{H}(\Delta z) = \sum_{i,j} J_{i,j} \hat{S}_i \cdot \hat{S}_j + D\hat{S}_{z,\text{Nc}}^2 + \sum_{i} J_{i,\text{Nc}}(\Delta z)\hat{S}_{\text{Nc}} \cdot \hat{S}_i$$

**Figure 1. Schematic representation of the spin model.** A poly-radical molecule is fitted to a spin model which is further coupled to the S=1 site modelling the NiCp$_2$ tip, illustrated schematically by the four arrows. Colour scale: grey – C, white – H, green – Ni, black – metallic atoms of STM tip. Blue and red lobes represent spin density of the molecule calculated by multireference CASCI method. Inset: Tipheight (Δz) dependent d$^2$I/dV$^2$ spectra, experimentally acquired using a NiCp2-functionalized tip, that are characteristic of the molecule's magnetic ground state.

In this work, we demonstrate the potential for NiCp₂-functionalized tips as spin-sensitive probes toward the precise discrimination of magnetic ground states of strongly correlated polyradical molecules, where the aforementioned methods often struggle to resolve subtle differences between spin configurations. It also enables us to map the inhomogeneous spatial distribution of the local exchange interaction of the polyradical π-magnet acting on an external spin. We show these on three different polyradical nanographenes: diradical **D1** (see Fig. 2a) and two isomeric triradicals **D2a, D2b** (see Figs. 3a,g)).

Using scanning tunneling spectroscopy (STS) and IETS acquired with NiCp₂-functionalized tips supported by the theoretical analysis, we can not only distinguish different spin multiplets but also resolve the number of unpaired spins, demonstrating a distinct response between the diradical **D1** and the triradicals **D2a, D2b**. More importantly, this method allows us to resolve the nearly degenerate ground states of the triradical dimers **D2**, confirming for **D2a** and **D2b,** respectively, the triradical doublet and quartet as their ground states. This discrimination highlights the ability of NiCp₂-functionalized tips to resolve the spatial distribution of spins and offers a clear advantage for spin mapping at the molecular level. The experimental findings are complemented by many-body calculations employing the complete active space configuration interaction (CASCI) method.

To rationalize the variation of the experimental IETS signal, we combine Heisenberg spin models and cotunneling theory[33] to simulate the corresponding inelastic tunneling current.[28] The Heisenberg model describes the tip-sample height-dependent exchange interaction $J_{Nc,i}(z)$ between a set of molecular ½-spins $\hat{S}_i$ and spin $\hat{S}_{Nc} = 1$ of the NiCp2 tip, schematically shown in Figure 1. The spin Hamiltonian $\hat{H}_{Nc}$ of NiCp2 tip is given by the out-of-plane anisotropy $\hat{H}_{Nc} = D\hat{S}_{Nc,z}^2$, with $D$ = 4 meV. The spin Hamiltonian of the molecule is described by a set of spin ½ sites: $\hat{H}_M = \sum_{i,j} J_{i,j} \hat{S}_i \cdot \hat{S}_j$, where the coefficients $J_{i,j}$ are chosen to reproduce the energy spectrum of the molecule obtained from many-body CAS calculations. Lastly, we include the tip-sample dependent exchange interaction $J_{Nc,i}$ between the local molecular spins $\hat{S}_i$ and the NiCp2 $\hat{S}_{Nc}$, $\hat{H}_{int}(z) = \sum_i J_{Nc,i}(\Delta z)\hat{S}_{Nc} \cdot \hat{S}_i$. Here Δz represents the variation of the tip-sample distance with respect to the closest distance. The exchange interaction between NiCp2 spin and local spins is defined as $J_{Nc,i}(\Delta z) = J_{0,i}\exp(-\lambda\Delta z)$. We can now write the full Heisenberg spin model as:

$$\hat{H}_{M,Nc}(\Delta z) = \hat{H}_M + \hat{H}_{Nc} + \hat{H}_{int}(\Delta z) = \sum_{i,j} J_{i,j} \hat{S}_i \cdot \hat{S}_j + D\hat{S}_{Nc,z}^2 + \sum_i J_{Nc,i}(\Delta z)\hat{S}_{Nc} \cdot \hat{S}_i . \quad (1)$$

However, the eigenvalues of the Heisenberg spin model (1) cannot be compared directly to the experimental IETS signal, as the intensity of IETS peaks corresponding to given spin excitations may strongly alternate. To account for the relative intensity of the different eigenstates of the spin model, we resort to cotunneling theory to calculate the inelastic current through the spin system (1) coupled to reservoirs.[28]

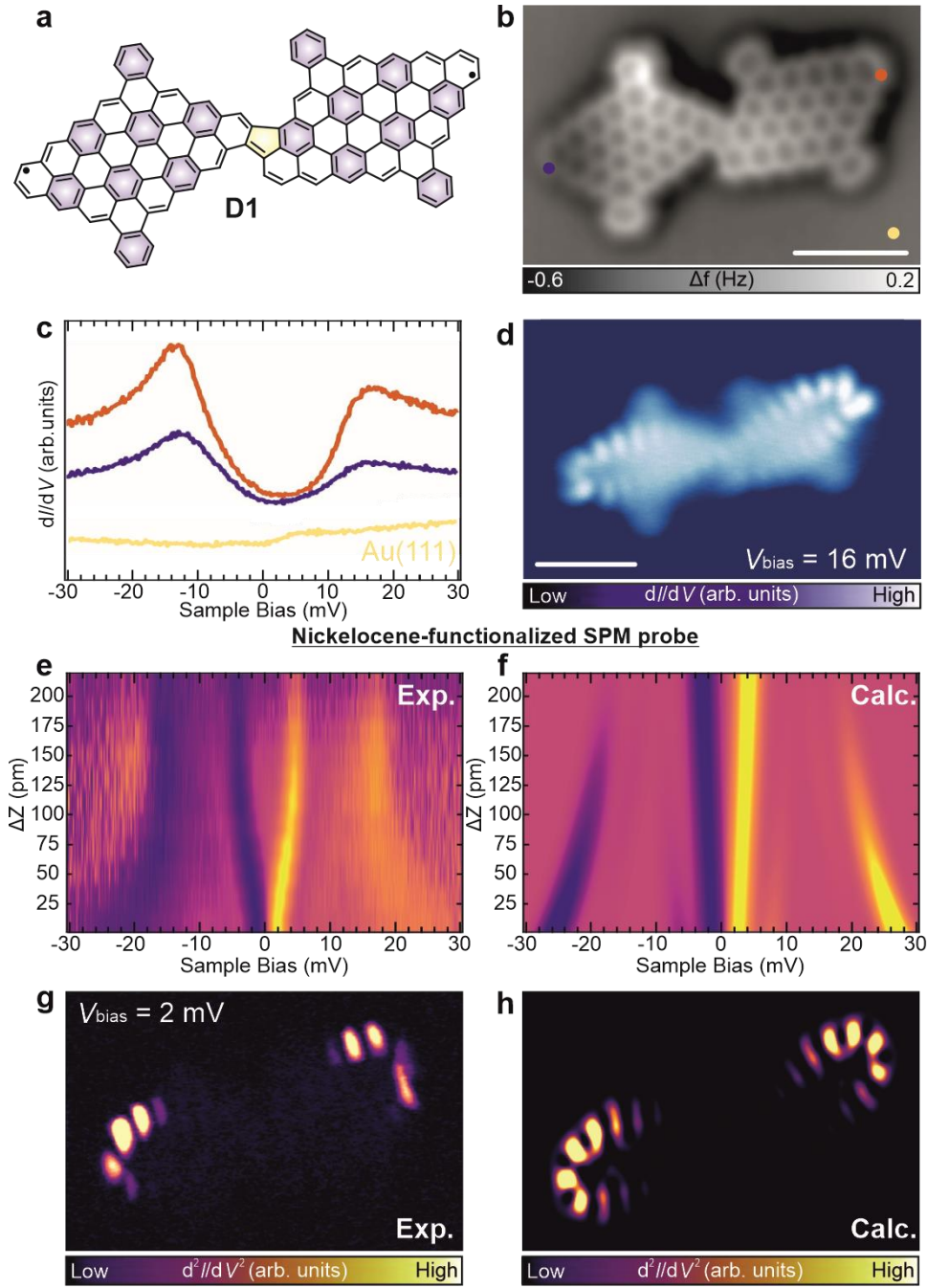

**Figure 2. Structural and electronic characterization of D1 on Au(111).** a) Chemical sketch of the dimer **D1** composed of two **I** NGs linked through a pentagon following a dehydrogenation reaction. b) Experimental constant height nc-AFM image acquired with CO-tip confirming the structure of **D1**. Scale bar = 1 nm. c) d$I$/d$V$ spectra of **D1** acquired at the positions marked with orange and blue circles in (a). From both spectra, we infer J$_{eff}$ ∼ 12meV. The yellow circle corresponds to the reference spectrum acquired on the bare Au(111) substrate. Spectra normalized and offset for clarity. Setpoints: $V_b$ = 30 mV, $I_t$ = 45 pA (orange, blue), $V_b$ = 30 mV, $I_t$ = 100 pA (yellow). d) Constant-height d$I$/d$V$ map of **D1** acquired at $V_b$ = 16 mV. Scale bar: 1 nm. e) Height-dependent map composed of a series of eleven d$^2I$/d$V^2$ spectra acquired with a NiCp$_2$ terminated scanning-probe tip on **D1** edges f) Theoretical simulation of the d$^2I$/d$V^2$ spectra corresponding to the experiment shown in (e). g) Constant-height d$^2I$/d$V^2$ map of **D1** acquired with a NiCp$_2$-functionalized tip at $V_b$ = 2 mV. h) Simulated constant-height d$^2I$/d$V^2$ map obtained by computing the overlap of NiCp$_2$ orbitals with the molecular spin density.

For the experimental part, we followed an on-surface synthesis approach to fabricate the three polyradical NGs (a diradical, **D1,** and the two isomeric triradicals **D2a, D2b**) by depositing a suitable molecular precursor on the Au(111) surface under UHV conditions. Subsequent annealing at 250 °C induces the oxidative ring closure and dehydrogenation reactions, together with a intermolecular coupling, which provides various nanographene products (for more details, see Fig. S17).*

We elucidated their chemical structure by means of high-resolution non-contact atomic force microscopy (nc-AFM) with a carbon monoxide (CO)-functionalized tip,[34] see Fig 2b and Fig 3b,h. Ab initio CASCI calculations predict that **D1** is a diradical with a singlet open-shell ground state and an excited triplet state at 12 meV. The calculations assign **D2a, D2b** as triradicals with nearly degenerate ground states, with the quartet and doublet states being separated by 4 meV. Such a small difference in energy makes it impossible to reliably assign the ground state using CASCI.

Next, we performed STS measurements of **D1** product (see Figs. 2a,b) to analyze its electronic structure. An indirect indication of the magnetic ground state of **D1** is manifested in the d$I$/d$V$ spectra acquired in the vicinity of the Fermi level. Fig. 2c shows such a d$I$/d$V$ spectrum featuring two conductance steps at 12 meV symmetrically positioned around the Fermi energy, which we tentatively assign then to inelastic spin-flip excitations from the ground state to the first excited magnetic states with excitation energy of J$_{eff}$ = 12 meV. This value matches well to the calculated energy gap between the singlet ground state and the first excited triplet state by CASCI(12,12) calculations, which provides a singlet-triplet gap of 12 meV (see Fig. S1 and Fig. S2 in the SOM for a description of the employed active space and the predicted diradical character). In addition, the constant-current d$I$/d$V$ maps obtained close to the spin excitation thresholds are shown in Fig. 2d. They match verywell with the calculated d$I$/d$V$ maps of Natural Transition Orbitals (NTO)[35] corresponding to the singlet-triplet transition obtained from CASCI calculations, see Figure S3. While there is good agreement between the experimental and theoretical d$I$/d$V$ maps and the inelastic energy gap (see Fig. 3c for the simulated map of the spin excitation), direct experimental evidence of the presence of the singlet ground state and first excited triplet state is missing.

Figure 2e shows z-dependent d$^2I$/d$V^2$ spectra acquired with NiCp$_2$-probe at edge of the **D1** (at orange marker in Figure 2b), which reveals significant variation of the d$^2I$/d$V^2$ signal upon NiCp$_2$-probe approach. In far distance, we observe two characteristic d$^2I$/d$V^2$ peaks at 4 meV and 16 meV at each polarity. These two peaks correspond, respectively, to the bare excitation of the NiCp$_2$ and the joint excitation of the molecule and the NiCp$_2$ (12+4=16 meV). The bare excitation of the molecule, with the NiCp$_2$ in its ground state, does not contribute to the current, as reproduced in the simulated map in Fig. 2f (see SOM for a detailed discussion). As the NiCp$_2$-probe approaches, we observe a gradual inward renormalization of the peak at 4 meV towards lower values, while the peak at 16 meV becomes broader and shifts to higher energies.

To rationalize the experimental d$^2I$/d$V^2$ spectra, we carried IETS simulations using the aforementioned theoretical model combining transport cotunneling theory with a Heisenberg model. Here, the molecule ($\hat{H}_M$ in Eq. 1 above) is represented by two-site two spins $\hat{S}_i = 1/2$ model $\hat{H}_M = J\hat{S}_1 \cdot \hat{S}_2$ where we set the parameter $J$ = 12 meV to reproduce the singlet ground state and the singlet-triplet energy gap 12 meV obtained from CAS calculations.

Figure 2f shows simulated z-dependent d$^2I$/d$V^2$ spectra which reproduce well the experimental data set. As the tip approaches, the exchange interaction increases and thus the singlet and triplet molecular states become mixed. Therefore, the spin of the molecule cannot be used as a quantum number, since its expected value deviates largely from the eigenvalues 0 and 1. This mixing of spin states at close tip-sample distances provides the characteristic renormalization corresponding to the particular singlet-triplet spin configuration of the molecule (see Fig. S13 and accompanying text for a discussion of the eigenstates of the Heisenberg model and their associated quantum numbers). It is worth noting that such a renormalization is characteristic of the antiferromagnetic dimer (being absent in the case of ferromagnetic coupling, see Figure S12 and underlying discussion in SOM) and is also due to the asymmetric coupling of the spin of the NiCp2-probe to one of the spins $\hat{S}_i$ of the molecular dimer: That is, in eq. (1) we set as $J_{Nc,1}(\Delta z) = J_{0,1} \exp(-\Delta \lambda z)$ and $J_{Nc,2}(\Delta z) = 0$. We set $J_{0,1} = 8\ meV$ and $\lambda = 1.5$ to fit the observed branching in the experimental range of 2 Å. In this way, we account for the local nature of the spin interaction. If the spin of NiCp2-probe coupled to the whole spin of the molecule, the inward renormalization of the bare excitation of NiCp2 at 4 meV would not be observed (see Fig, S15 and the accompanying text in the SOM for details). Also the variation of $d^2I/dV^2(z)$ signal is strongly site dependent. For example in the center of the molecule, the $d^2I/dV^2(z)$ signal remains constant upon the NiCp2-probe approach. These observations indicate strongly inhomogenous distribution of magnetic signal over the molecule, due to presence of spatially extended unpaired π-electrons.

Thus we recorded spatial IETS maps at 2 meV, to capture spatial distribution of the variation of the bare NiCp2 excitation. Fig. 2g reveals a localization of the IETS signal predominantly at the two edges, which nicely coincides with calculated spin-density distribution shown in Figure S2. To rationalize the experimental map, we calculated spatial NiCp2 maps for dimer **D1** using a Heisenberg Hamiltonian that incorporates spatially dependent exchange interactions between the NiCp2 tip and the molecular spin centers. The spatial exchange interactions are derived from the wavefunction overlap between the NiCp2 and molecular spin density, following Chen's derivative rule[36]. Using this Hamiltonian, the spatially resolved inelastic current is calculated as a function of the bias voltage, where the NiCp2 excitation undergoes renormalization. The very good match between the experimental and calculated NiCp2 spatial maps, compare Figs. 2g and h, demonstrate that the NiCp2 spatial $dI/dV$ maps provide an accurate representation of the molecule's spin-density distribution, in good agreement with the radical character predicted from an ab initio analysis (see the Natural Orbitals in Fig. S2 in the SOM).

Next, we focus on the electronic and magnetic properties of dimers **D2a** and **D2b**. These two dimers are structural isomers that differ only by a reflection on one-half of the molecule (see Figs. 3A,G). Experimental $dI/dV$ spectra show, in both cases, the coexistence of a Kondo peak on the right wing of the dimers and a spin excitation on the left wing (see F
igs. 3C,I). Ab initio CASCI(11,11) with NEVPT2 corrections yield for the two isomers identical electronic properties: Both dimers **D2** are predicted to be fully triradical molecules with a doublet ground state and a quartet ground state at $J_{eff}$ =4 meV, with the next excited state (another triradical doublet) around 0.5 eV higher in energy (see Figs. S4 and S5 in the SOM for the orbitals forming the active space). However, due to the small spectral gap between the ground and the first excited state, the theoretical calculations do not allow for a conclusive determination of the ground state.

Thus, we tried to discern the ground state using the spatial distribution of low-energy $dI/dV$ maps corresponding to spin-excitation and Kondo resonance. To rationalize the experimental STS maps shown in Fig. S18, we calculated NTO and Kondo orbitals (KO) for both doublet and quartet ground states,[37] as described in detail in SOM. The resulting theoretical $dI/dV$ maps of NTO and KO (see Figs S8-11) for the doublet and quartet ground state are very similar, and they match the experimental evidence very well. Therefore, the spatial distributions of the spin excitation and Kondo signal are also insufficient to determine whether the triradical ground states of **D2a** and **D2b** are doublets or quartets.

We then focus on the d$^2I$/d$V^2$ spectra acquired with NiCp$_2$-probe on the left wing of the dimers **D2**, where the spin excitation is visible. As shown in Figs.3E,K, these spectra exhibit distinct features for each structural isomer; **D2a** displays curved branches originating from the renormalized excitation lines of the NiCp$_2$ excitation, while **D2b** shows two parallel lines with features on the upper side of the map that are not visible. These are robust indicators of distinct ground states: a triradical doublet and a quartet, respectively. Natural Orbitals from CASCI calculations show the presence of one radical on the left wing (where the spin excitation is visible) and two radicals on the right wing (where the Kondo signal is visible), see Figs. S6 and S7 in the SOM for details. The location of these three radicals is schematically shown in Figs. 3D,J. To model these spectra, we resort to a Heisenberg spin model following equation 1 with three spin sites (see Figs. S14 and S15 for the local basis of orbitals on which the spin model is defined).

In this case, however, we construct three-sites molecular spin Hamiltonians, $\hat{H}_M$, one that has the doublet ground state, $\hat{H}_{M,d}$, and one that has the quartet ground state, $\hat{H}_{M,q}$ (see Figs. 3D,J). These Hamiltonians are respectively given by $\hat{H}_{M,d} = J_d \hat{S}_1 \cdot \hat{S}_2 + J_d \hat{S}_2 \cdot \hat{S}_3 + J_t \hat{S}_1 \cdot \hat{S}_3$ and $\hat{H}_{M,q} = J_q \hat{S}_1 \cdot \hat{S}_2 + J_q \hat{S}_2 \cdot \hat{S}_3 + J_t \hat{S}_1 \cdot \hat{S}_3$. The coefficients are fitted to reproduce the experimentally obtained spin excitation for, respectively, **D2a** and **D2b** (see Figs. 3D,J). We set $J_d = 4\ meV$, $J_q = -2\ meV$ and $J_t = -200\ meV$ to reproduce the spin correlations of the ab initio CAS calculations (see SOM for details). The large ferromagnetic interaction on the right-side radicals ensures that the second doublet excited state is far away, as indicated by the Ab initio CASCI(11,11) calculations. Therefore, Hamiltonian $\hat{H}_{M,d}$ has a doublet ground state and a quartet excited state at 4 meV, while Hamiltonian $\hat{H}_{M,q}$ has a quartet ground state and an excited doublet state at 2 meV. We now plug these Hamiltonians in eq. 1 with $J_{Nc,2}(z) = J_{0,2} \exp(-\lambda \Delta z)$ and $J_{Nc,1}(z) = J_{Nc,3}(z) = 0$ to account for the local coupling of the NiCp$_2$-probe with the single radical on the left wing of the dimers **D2** (see Figs. 3D,J). The resulting simulated d$^2I$/d$V^2$ spectra for the doublet and quartet model (Figs 3L,J) match well with the experimentally maps shown in Figs. 3E and K. This nice agreement allows us to conclude that the dimers **D2a** and **D2b** have doublet and quartet ground states, respectively. It is important to highlight that the spectroscopic characteristics of the maps remain robust despite perturbations of the spin excitation: the behaviour of the central branches remains unaltered for a range of different spin excitations (see Fig. S16 in SOM). This is crucial to conclude that the NiCp$_2$ measurements allow us to discriminate unambiguously between different ground states of a single molecule.

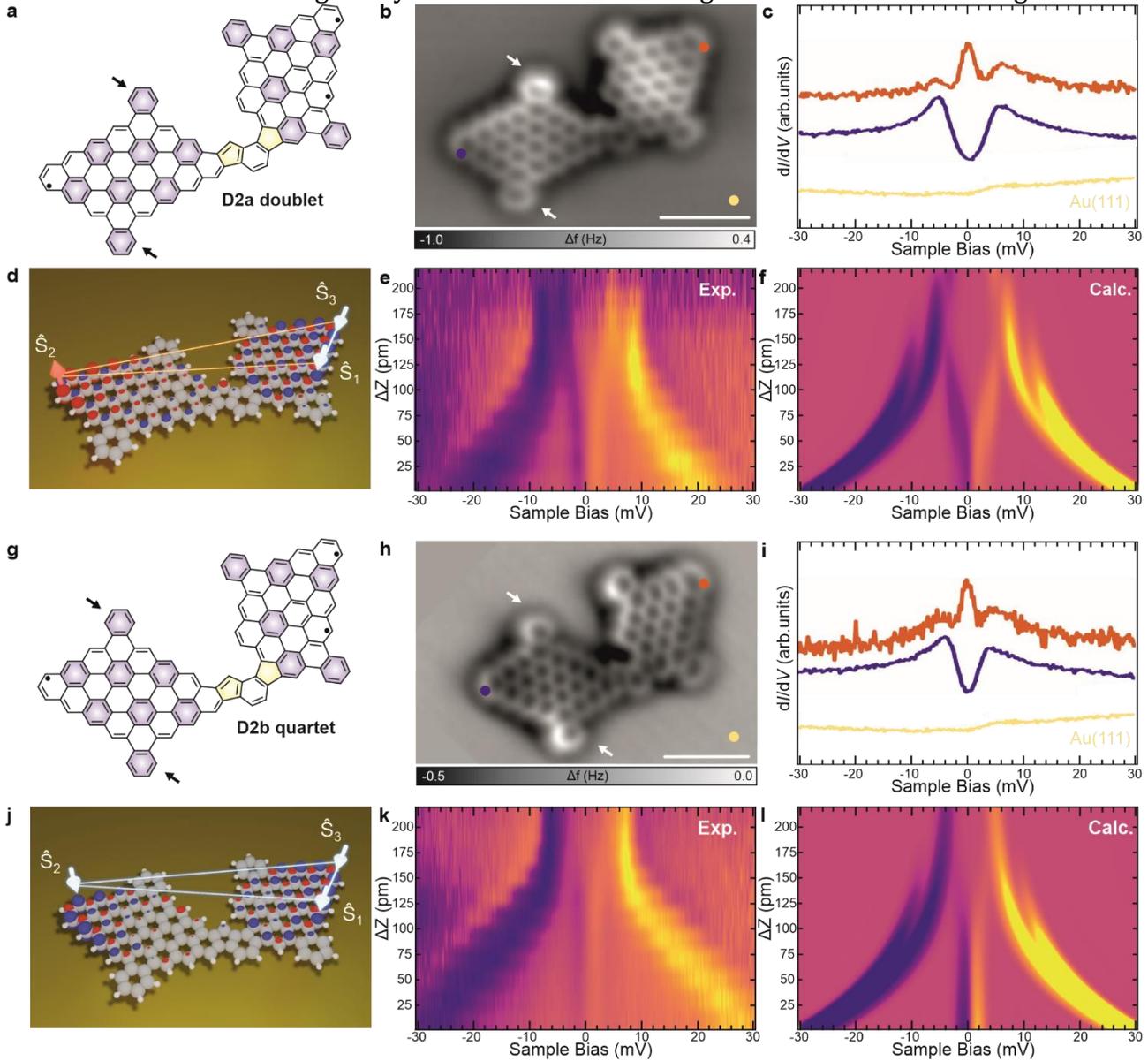

**Figure 3. Structural and electronic characterization of dimers D2 on Au(111).** a),g) Chemical sketches of dimers **D2a** and **D2b**. b),h) Constant height nc-AFM images acquired with a CO-tip confirm the structure of dimers **D2a** and **D2b**. Scale bars: 1 nm. c),i) STS measurements acquired at the positions indicated in b) and h). Spectra normalized and offset for clarity. Setpoints: c) $V_b = 30$ mV, $I_t = 110$ pA (purple), $I_t = 60$ pA (orange), $I_t = 100$ pA (gold). i) $V_b = 30$ mV, $I_t = 60$ pA (purple), $I_t = 30$ pA (orange), $I_t = 100$ pA (gold). d),j) Schematic spin models with doublet and quartet ground states employed to model Dimer **D2a** and **D2b**, respectively. In both cases spins $\hat{S}_1$ and $\hat{S}_3$ are ferromagnetically coupled. $\hat{S}_2$ is antiferromagnetically coupled to $\hat{S}_1$ and $\hat{S}_3$ in the case of the doublet (d) and ferromagnetically coupled to $\hat{S}_1$ and $\hat{S}_3$ in the case of the quartet (j). Blue/red lobes indicate spin density calculated by the CASCI method. e),k) Tip-height dependent maps composed of a series of eleven $d^2I/dV^2$ spectra acquired with a NiCp2 terminated scanning-probe tip on the left wings (purple markers in b, h) of **D2a** and **D2b** at different tip-sample distances (20 pm between spectra). f),l) Theoretical simulations of the $d^2I/dV^2$ spectra corresponding to the experiments shown in e) and k), obtained for the doublet and quartet ground state sketched in d) and j), respectively.

In conclusion, the particular spin state of a single molecular magnet, due to the exchange interaction with the nickelocene functionalized probe, generates a unique IETS signal response. This tip-height dependent IETS signal characteristic for a particular spin state allows for the discrimination of magnetic ground states of a given molecular magnet. This method significantly expands the possibilities of the SPM technique for characterizing the magnetic properties of individual molecules. At the same time, it opens new possibilities for the study of complex spin systems such as π-d or spin-frustrated molecular systems with high spatial resolution. In principle, this method can also be used for 3D mapping of the exchange field of molecular magnets. We anticipate that this method can be extended by incorporating higher-order scattering terms, enhancing its sensitivity toward complex molecular spin states.

## Methods

### CAS calculations

First, the geometry of free-standing molecules is optimized with the DFT code FHI-AIMS code[38] with the PBE0 functional[39]. As the molecules are strongly polyradical, we have employed the many-body CASCI method for a precise description of their electronic structure, resolving the full many-body Hamiltonian:

$$\widehat{\mathcal{H}}_{\text{CAS}} = \sum_{i,j,\sigma} t_{ij}\, \hat{C}^{\dagger}_{i\sigma}\hat{C}_{j\sigma} + \sum_{i,j,k,l,\sigma,\sigma'} \mathcal{V}_{ijkl}\, \hat{C}^{\dagger}_{i\sigma}\hat{C}^{\dagger}_{j\sigma'}\hat{C}_{k\sigma'}\hat{C}_{l\sigma},$$

where $\hat{C}^{\dagger}_{i\sigma}(\hat{C}_{j\sigma})$ are the creation (annihilation) operators are associated with the basis of molecular orbitals. We have calculated the one and two-body integrals $t_{ij}$ and $\mathcal{V}_{ijkl}$ from the quantum chemistry software ORCA.[40]

The orbitals of the active space are obtained from DFT with PBE functional[41]. We employ closed-shell DFT orbitals for the case of an even number of electrons (dimer **D1**) and restricted open-shell (ROKS) DFT orbitals for the case of an odd number of electrons (dimers **D2a** and **D2b**). For further correction of the out-of-CAS dynamical electron correlation, we employed the Quasidegenerate Second-Order N-Electron Valence State Perturbation Theory (QD-NEVPT2).[42]

### Natural Orbitals

We construct the spinless 1-particle reduced density matrix from the wavefunction obtained from our CASCI calculation.[43] The diagonalization of the density matrix provides the natural orbitals, and their occupations yield the number of unpaired electrons.

### Natural Transition Orbitals

To rationalize the d$I$/d$V$ maps of the spin excitation, we employ the so-called Natural Transition Orbitals (NTO) for the spin-flip operator[35] encoding the transition between the ground $\Psi_0$ and excited $\Psi_1$ states. NTO orbitals are obtained from the diagonalization of the matrix $TT^\dagger$, where the matrix $T$ is given by elements

$$T_{jk} = \langle \Psi_1 | \hat{C}_{j\uparrow}^\dagger \hat{C}_{k\downarrow} | \Psi_0 \rangle,$$

where the indices $j, k$ run over the active orbitals.

**Kondo Orbitals**

To account for the spatial distribution of the Kondo signals, we employ the formalism of Kondo orbitals.[37] Here, we diagonalize the coupling matrix obtained from a multi-level Anderson Hamiltonian, which models the coupling of the molecules (described by the CASCI Hamiltonian) to the metallic substrate. This allows us to identify the antiferromagnetic channels contributing to the scattering process that account for the Kondo screening mechanism.

**Simulation of d$I$/d$V$ maps** Theoretical d$I$/d$V$ maps of NTOs and Kondo orbitals were calculated by the Probe Particle Scanning Probe Microscopy (PP-SPM) code[44] for a CO-like tip, which was represented by a linear combination of P$_x$P$_y$ (85%) and s-like (15%) orbitals without tip relaxation.

**Spin models**

We fit dimer **D1** to a two-site spin model given by:

$$\hat{H}_{D1} = J_1 \vec{\hat{S}}_1 \cdot \vec{\hat{S}}_2,$$

with $J_1 = 12$ meV. Dimers **D2** are fitted to a three-site spin model, where site 2 corresponds to the lone radical on the left wing (see Figs. S8b,S9b), and sites 1 and 3 correspond to the two radicals on the right wing of the molecule (see Figs. S8b,c,S9b,c). The fitting can be done to reproduce either a doublet or a quartet ground state:

$$\hat{H}_{D2,d} = J_t \vec{\hat{S}}_1 \cdot \vec{\hat{S}}_3 + J_d \left( \vec{\hat{S}}_1 + \vec{\hat{S}}_3 \right) \cdot \vec{\hat{S}}_2$$
$$\hat{H}_{D2,q} = J_t \vec{\hat{S}}_1 \cdot \vec{\hat{S}}_3 + J_q \left( \vec{\hat{S}}_1 + \vec{\hat{S}}_3 \right) \cdot \vec{\hat{S}}_2,$$

with the coefficients for the spin models as specified in the main text.

Calling in general $\hat{H}_M$ to any of the molecular spin Hamiltonians $\hat{H}_{D1}$, $\hat{H}_{D2,d}$ or $\hat{H}_{D2,q}$ above, our goal is now to connect such molecular Hamiltonian $\hat{H}_M$ to Hamiltonian $\hat{H}_{Nc}$ representing spin of NiCp2 tip, which is described by S=1 spin site with an out-of-plane magnetic anisotropy of 4 meV, $\hat{H}_{Nc} = D\hat{S}_{Nc,z}^2$.

To complete the model, we need to specify how the spin on the Nickelocene tip is coupled to the spins of nanographene through the interaction Hamiltonian $\hat{H}_{int}(\Delta z)$, where $\Delta z$ represents the variation in the tip-sample distance with respect to the closest distance ($\Delta z = 0$). We assume an antiferromagnetic interaction that arises from the kinetic exchange. The interaction Hamiltonian $\hat{H}_{int}$ has then the form

$$\hat{H}_{int}(\Delta z) = \sum_i J_{Nc,i}(\Delta z) \hat{S}_{Nc} \cdot \hat{S}_i,$$

where the couplings $J_{Nc,i}(\Delta z)$ will decay exponentially with height, $J_{Nc,i}(\Delta z) = J_{0,i} \exp(-\lambda \Delta z)$. However, due to the localization of the radicals on the nano-graphenes (see Fig. S2 and Fig. S6, S7), the spin of the Nickelocene couples only to one spin site of the model. The total spin Hamiltonian reads as:

$$\hat{H}_{M,\text{Nc}}(\Delta z) = \hat{H}_M + \hat{H}_{\text{Nc}} + \hat{H}_{\text{int}}(\Delta z).$$

**Cotunneling Theory**

To calculate the $d^2I/dV^2$ maps from the spin models, we employ the perturbative approach described in reference,[33] as implemented for the case of NiCp2 tips and molecular systems.[28] The cotunneling results from a second-order perturbation calculation starting from the Hamiltonian:

$$\hat{H}(z) = \hat{H}_{M,\text{Nc}}(\Delta z) + \hat{H}_{\text{tip}} + \hat{H}_{\text{sub}} + \hat{H}_{\text{tun}}(\Delta z),$$

where $\hat{H}_{M,\text{Nc}}$ is given above, $\hat{H}_\eta = \sum_{k,\sigma} \varepsilon_{\eta,k,\sigma} \hat{a}^\dagger_{\eta,k,\sigma} \hat{a}_{\eta,k,\sigma}$, with $\eta = $ tip,sub, and $\hat{H}_{\text{tun}}$ is the tunneling Hamiltonian given by

$$\hat{H}_{\text{tun}}(\Delta z) = \sum_{i,\sigma,\sigma'} T_{0,k,k'} \left( \hat{a}^\dagger_{\text{tip},k,\sigma} \hat{a}_{\text{sub},k,\sigma'} + \text{h.c.} \right) +$$

$$\sum_{i,\sigma,\sigma'} T_{i,k,k'}(z) \vec{P}_{\sigma,\sigma'} \cdot \vec{S}_i \left( \hat{a}^\dagger_{\text{tip},k,\sigma} \hat{a}_{\text{sub},k,\sigma'} + \text{h.c.} \right),$$

where $i$ labels the sites of the spin Hamiltonian $\hat{H}_{M,\text{Nc}}$, $\vec{P}$ are the $\frac{1}{2}$ Pauli matrices and $T_{i,k,k'}(\Delta z)$ are the coupling between electrons and spins, which include a dependence on the tip-sample distance $z$. We take the couplings with the molecular sites to decay exponentially with the height, $z$, as $T_{i,k,k'}(\Delta z) = T_i(\Delta z) = T_{i,0} \exp(-\mu \Delta z)$. In our simulations, we fix $\mu_i$ and the $T_{i,0}$ so that $T_{\text{Nc}}(\Delta z) = 3$ for all $z$ and $T_i(z)$ has value 3 at a closest distance and decays exponentially to 1 at the furthest distance (2 Å above). The current is then computed in second order as :

$$I_{inel}(V) = \sum_{\alpha,\alpha',\eta=x,y,z} P_\alpha \left| \langle \alpha | \sum_{i \in \text{site}} T_i(\Delta z) \hat{S}_{i,\eta} | \alpha' \rangle \right|^2 F_{\alpha,\alpha'}(V),$$

where

$$F_{\alpha,\beta}(V) = \left( \frac{V - E_\alpha + E_{\alpha'}}{1 - \exp(-\beta(V - E_\alpha + E_{\alpha'}))} + \frac{V + E_\alpha - E_{\alpha'}}{1 - \exp(\beta(V + E_\alpha - E_{\alpha'}))} \right),$$

and $P_\alpha$ are the populations of the eigenstates $\alpha$ at thermal equilibrium $P_\alpha = \exp(-\beta E_\alpha)$.

**Sample preparation**

The Au(111) crystal was prepared by several cycles of sputtering with Ar$^+$ ions and subsequent annealing at ~500 °C in ultra-high vacuum (UHV). The precursor molecules, 6,14-Bis(2,6-dimethylphenyl)-3,11-diphenyldibenzo[*hi,st*]ovalene (DBOV-Ph), were sublimed into the UHV chamber from a Knudsen cell via heating at 400 °C for 10 minutes, resulting in a coverage of ~0.3 ML. The Au(111) surface was held at room temperature during deposition. Subsequently, the sample was annealed at ~250 °C for 12 minutes to facilitate both the formation of planar molecules via cyclodehydrenation, and the oxidative ring closure reaction which resulted in the formation of the dimers **D1, D2a** and **D2b** (Fig. S17).

**SPM Measurements**

All scanning probe microscopy (SPM) measurements were performed in a SPECS-JT microscope at a measuring temperature of 4 K using a Kolibri sensor (f$_0$ ~ 1 MHz, Q ~160k, K ~1800 N/m, 50 pm amplitude modulation). STS, d$I$/d$V$ maps, and ncAFM measurements were all performed by first functionalizing the tip with a carbon-monoxide (CO) molecule. A lock-in method was used for both STS and d$I$/d$V$ maps using $V_{mod}$ = 1 mV, $f_{mod}$ = 723 Hz.

### NiCp$_2$ Measurements

Nickelocene (NiCp$_2$) molecules were deposited from a tantalum pocket kept at room temperature directly onto the sample in the microscope (T<10 K). The tip was functionalized with a NiCp$_2$ molecule by repeatedly scanning over an individual molecule on the Au(111) surface with $V_b$ < 4 mV, $I_t$ < 50 pA until spontaneous functionalization occurred. NiCp$_2$ tips were characterised by the sharpness of their imaging, by their stability in ncAFM signal as measured by forward and backward Δf(Δz) curves, and finally by the presence of inelastic signatures at ±4 mV on bare Au(111) (see Fig. S18 in the SOM). A lock-in method was used to record d$^2$I/dV$^2$ for acquiring both spectra (e.g. Figure 1) and maps (e.g. Figure 2g), with $V_{mod}$ = 1 mV, $f_{mod}$ = 723 Hz.

# Supporting Information

## Authors and Affiliations

### Authors

**Institute of Physics of the Czech Academy of Science, Praha162 00, Czech Republic**

Diego Soler-Polo, Oleksandr Stetsovych, Manish Kumar, Benjamin Lowe, Andrés Pinar Solé and Pavel Jelínek

**IMDEA Nanoscience, Madrid 28049, Spain**

Ana Barragán, Elena Pérez-Elvira, David Écija and José I. Urgel*

**Okinawa Institute of Science and Technology Graduate University, Kunigami-gun, Okinawa 904-0495, Japan**

Zhiqiang Gao, Hao Zhao, Goudappagouda and Akimitsu Narita

### Corresponding authors

Diego Soler-Polo, Akimitsu Narita, Pavel Jelínek and José I. Urgel

### Author Contributions

### Ethics declarations
Competing interests
The authors declare no competing interests.

### Acknowlegments

The authors acknowledge to several funding organizations for their financial support. We thank support from the TEC-2024/TEC-459-(SINMOLMAT-CM) and '(MAD2D-CM)-IMDEA-Nanociencia' projects funded by Comunidad de Madrid, by the Recovery, Transformation and Resilience Plan, and by NextGenerationEU from the European Union, and from the Spanish Ministry of Science, Innovation and Universities (Project PID2022-136961NB-I00 and PID2023-152793NA-I00). IMDEA Nanociencia also acknowledges the "Severo Ochoa" Programme for Centers of Excellence in R&D (MINECO, Grant SEV-2016-0686 and CEX2020-001039-S). We are grateful to the financial support from the Okinawa Institute of Science and Technology Graduate University, JSPS International Joint Research Program (JRP-LEAD with DFG) No. JPJSJRP20221607, and JSPS KAKENHI Grant No. JP22F22031 and JP23KF0075. H.Z. and G. acknowledges the JSPS Postdoctoral Fellowship for Research in Japan. We appreciate financial support from the CzechNanoLab Research Infrastructure supported by MEYS CR (LM2018110) and GACR 23-05486S. J.I.U. and A.B. acknowledge the funding from MCIU for the Ramón y Cajal (RYC2022-037352) and Juan de la Cierva (FJC2021-046524-I) programs, respectively. B.L. acknowledges support from GACR 25-16632I. This work was supported by European Structural and Investment Fundsand the Czech Ministry of Education, Youth and Sports(Project No TERAFIT - CZ.02.01.01/00/22_008/0004594). The authors also wish to thank Federico Frezza for valuable discussions.

## Refeerences